\documentclass[aps,reprint,superscriptaddress]{revtex4-2}
\usepackage{amssymb}
\usepackage{amsmath,bm}
\usepackage{graphicx,color}
\usepackage[dvipsnames]{xcolor}
\usepackage{bbm}
\usepackage[bottom]{footmisc}
\usepackage{multirow}
\usepackage{soul}

\usepackage[normalem]{ulem}
\usepackage{hyperref}
\hypersetup{
    pdftoolbar=true,
    pdfmenubar=true,
    bookmarksopen=true,
    bookmarksnumbered=true,
    breaklinks=true,
    linktocpage,
    colorlinks=true,
    linkcolor={blue!95!black},
    citecolor={teal!70!black},
    urlcolor={blue!95!black},
}

\begin{document}
\title{Microscopic Basis for Recovery Rheology and the Nonequilibrium Structure,Yielding, and Flow of Dense Particle Suspensions }

\author{Anoop Mutneja}
\email{anoop.mutneja2011@gmail.com} 
\affiliation{Department of Materials Science and Engineering, University of Illinois, Urbana, IL 61801, USA}
\affiliation{Materials Research Laboratory, University of Illinois, Urbana, IL, 61801, USA}
\author{Kenneth S. Schweizer}
\email{kschweiz@illinois.edu}
\affiliation{Department of Materials Science and Engineering, University of Illinois, Urbana, IL 61801, USA}
\affiliation{Materials Research Laboratory, University of Illinois, Urbana, IL, 61801, USA}
\affiliation{Department of  Chemistry, University of Illinois, Urbana, IL 61801, USA}
\affiliation{Department of  Chemical \& Biomolecular Engineering, University of Illinois, Urbana, IL 61801, USA}

\begin{abstract}{
The recent introduction of recovery rheology has provided qualitatively new physical insights into the yielding and flow of soft matter systems across diverse mechanically driven nonequilibrium protocols by separating the deformation strain into recoverable and unrecoverable components. A striking finding is that the fluid-like response associated with the gradually increasing unrecoverable strain ultimately leads to the continuous yielding transition from a solid to a liquid. We build on the force and particle level Elastically Collective Nonlinear Langevin Equation theory of activated dynamics within a nonequilibrium microrheological framework to formulate a general statistical mechanical foundation of step-rate start-up shear response that relates recovery rheology to microscopic structure, relaxation, and elasticity. Quantitative applications to metastable hard and soft sphere colloidal suspensions reveal testable new predictions and interconnections between macroscopic and microscopic properties: (i) the steady state recoverable strain is directly related to the steady-state shear thinning; (ii) the transient stress overshoot amplitude varies non-monotonically with packing fraction and is quantitatively linked to the steady-state recoverable strain; (iii) the acquired unrecoverable strain dictates the stress overshoot strain; (iv) the predicted enormous reduction of the structural relaxation time under deformation is inversely related to the unrecoverable strain-rate.
}
   
\end{abstract}

\maketitle

\section{Introduction}
 Dense kinetically arrested materials with liquid-like microstructure yet solid-like mechanical response are ubiquitous, encompassing diverse systems such as colloids, nanocomposites, amorphous metals, polymers, and cell tissue. Remarkably, such diverse materials can exhibit qualitatively similar responses to external shear deformation, albeit with largely system and thermodynamic state-specific quantitative aspects  \cite{bonn2017yield-35d,nicolas2018deformation-6dc,roberto2023introduction-384,karmakar2010statistical-b1a,leishangthem2017yielding-897,zausch2008from-c62}. Under strain-controlled step-rate deformation, the material initially elastically acquires stress, followed by an anelastic or ``strain softening” regime, then a stress overshoot indicating a continuous elastic-viscous crossover or yielding, and ultimately nonequilibrium steady state flow. This rich transient nonlinear mechanical response is of high practical importance. But its microscopic connection to the underlying coupled temporal evolution of nonequilibrium structure, elasticity, and stress and structural relaxation remains poorly understood. Thus, quantitative interconnections between observable macroscopic mechanical properties in quiescent, transient, and flowing states and internal microscopic material-specific properties remain largely a mystery. 

Modern iterative recovery rheology is providing much deeper insight into the underlying physics. Stress is measured as a function of time or accumulated total strain, $\gamma$, which is separated into recoverable, $\gamma_{rec}$, and unrecoverable, $\gamma_{unrec}$, components. The accumulation of unrecoverable strain drives a continuous material fluidization  \cite{shi2023benefits-879, singh2021revisiting-26e, kamani2021unification-578, donley2020elucidating-c91,keane2025universal-80b}, and holds promise to unify characterization of yielding phenomena across different nonlinear rheological measurements, including step-rate, step-strain, oscillatory shear, and nonlinear creep. The phenomenological continuum model by Kamani, Donley, and Rogers, the KDR model \cite{shi2023benefits-879, singh2021revisiting-26e, kamani2021unification-578, donley2020elucidating-c91,keane2025universal-80b} built on the recovery rheology perspective can reproduce many rheological features and trends based on 6 parameters determined from specific experimental measurements and data fitting. Importantly and rather remarkably, upon fine tuning of a specific and nonuniversal phenomenological parameter, the KDR model reproduces a qualitatively correct transient response under large amplitude oscillatory shear (LAOS) using the steady‑state material relaxation parameters and quiescent elastic parameters extracted from experiment. However, more broadly, a microscopic theoretical foundation for recovery rheology and the concept of recoverable versus unrecoverable strain remains absent, and creation of one is the primary goal of this article. At the same time, fundamental physical differences between the KDR model assumptions and our microscopic statistical mechanical theory are identified, suggesting new avenues for future experimental studies and perhaps refinement of the KDR framework. 

We employ a suite of statistical mechanical theory ideas formulated at the microscopic level of forces and particles, which have been extensively tested against experiment under quiescent and mechanically nonequilibrium conditions  \cite{schweizer2005derivation-c99,kobelev2005strain-2be,mirigian2014elastically-4c6,ghosh2023microscopic-39e,mutneja2025microscopic-f31,mei2025mediumrange-05d,chaki2024theoretical-92e,athanasiou2025probing-0e7}. The foundational framework is the Elastically Collective Nonlinear Langevin Equation (ECNLE) theory, which microscopically describes kinetic constraints and stochastic activated particle motion, and allows the prediction of macroscopic observables such as the activated relaxation time and diffusion constant, and the emergent elastic modulus of the transiently localized dynamical state. Under external deformation, these microscopic quantities evolve in a predictable manner, thereby enabling the theory to address emergent nonequilibrium variables critical to measurable macroscopic responses, including the acquired mechanical stress and a scalar order parameter that quantifies deformation‑modified structure. The essential new physical idea discussed here is to connect this microscopic structural deformation parameter to the macroscopic recoverable strain, and use this encoded deformation memory via the stress and recoverable strain to self-consistently predict nonequilibrium activated relaxation and elasticity at a given time/strain, and vice versa.

As a preview of our results, the following new insights and testable interconnections have been obtained for metastable glass-forming Brownian hard particle fluids and colloidal suspensions. A) Steady state shear thinning of the viscosity and structural alpha relaxation time ($\tau_\alpha$) as a function of shear rate, $\dot{\gamma}$, obeys an effective power law $\tau_{\alpha,SS}\left(\ \dot{\gamma},\phi\right)\sim{\dot{\gamma}}^{-\left(1-a\left(\phi\right)\right)}$ with a packing fraction ($\phi$) dependent exponent, $a\left(\phi\right)$. The systematic deviations from the ideal shear thinning behavior ($a=0$) that have been observed in experiment are microscopically understood to arise from a fundamental link to the nonuniversal growth of accumulated recoverable strain with shear rate, $\gamma_{rec,SS}\sim{\dot{\gamma}}^{a\left(\phi\right)}$. B) The \textit{transient} stress overshoot amplitude is determined by the \textit{steady state} recoverable strain, which quantifies nonequilibrium structural deformation, implying a deep connection between transient and long-time nonequilibrium physics. C) The corresponding overshoot strain is determined by a nonuniversal critical level of unrecoverable strain, thus revealing that the active deformation-driven solid-to-fluid crossover is tied to the accumulation of the unrecoverable strain component of the total macroscopic strain. D) The difficult to experimentally measure structural relaxation time under active deformation is predicted to be obtainable from the macroscopic unrecoverable shear rate, allowing a link to a key phenomenological ansatz employed in the KDR model. 

\section{Theory}
\begin{figure*}
	\includegraphics[width=0.9\textwidth]{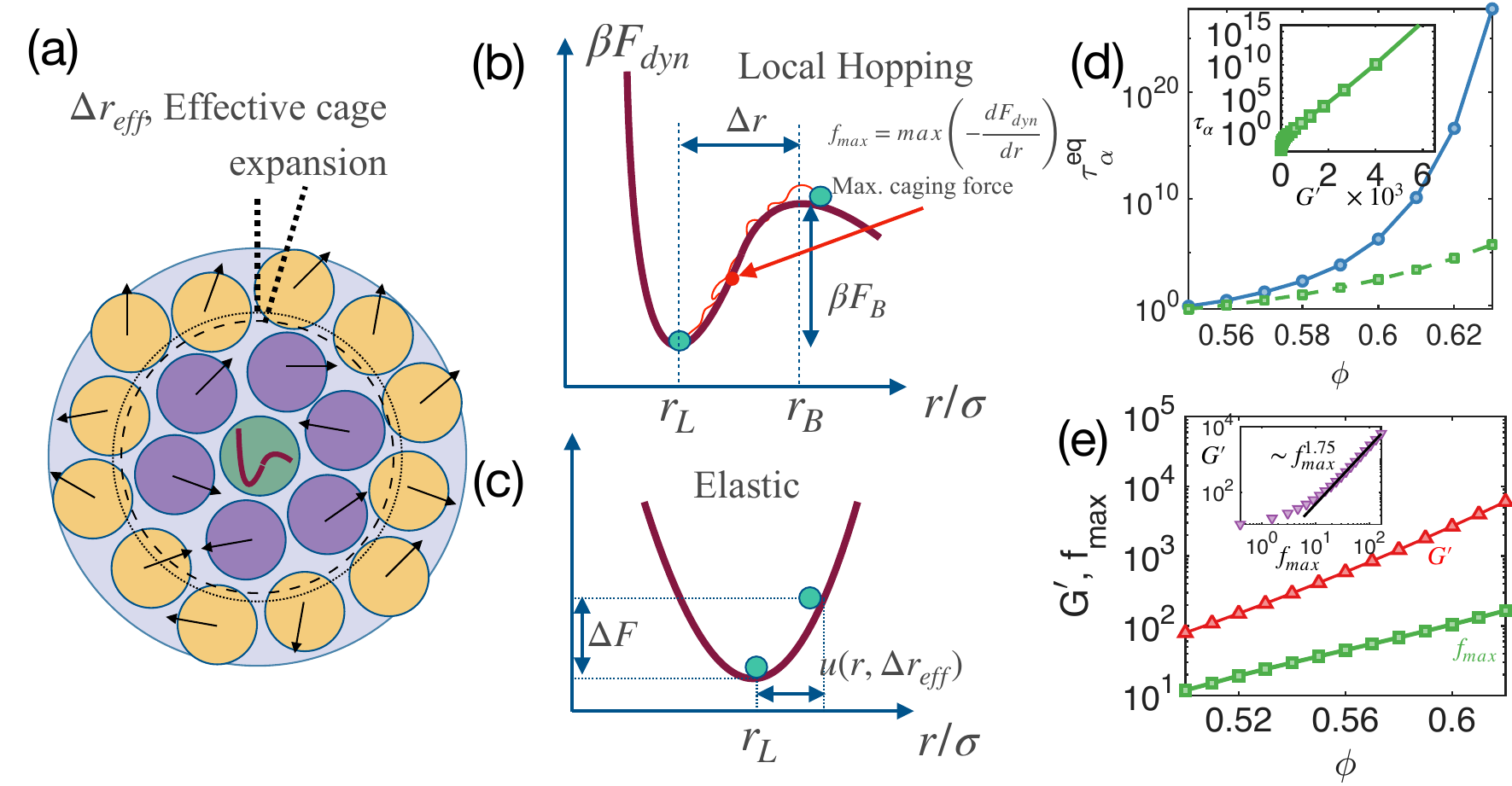}
	\caption{\textbf{Quiescent ECNLE Theory Background Elements } \cite{schweizer2005derivation-c99,mirigian2014elastically-4c6,mei2025mediumrange-05d} : (a) Graphic representation of the ideas of ECNLE theory where a central tagged particle (green) is transiently localized in a spatially-resolved confining caging potential formed by its identical neighbours (purple), along with the longer range (yellow) small collective elastic displacements of all particles outside the cage region (arrows) required to causally facilitate a cage escape event. The net structural alpha relaxation event is of a  spatially local-nonlocal character. (b) The confining dynamic free energy as a function of tagged particle displacement in units of particle diameter ($\sigma$) with important length and energy scales indicated where $\beta={{(k}_BT)}^{-1}$. The maximum caging force $f_{max}$ (in units of $k_BT\sigma^{-1})$ experienced by a tagged particle. (c) Harmonic collective elastic displacement of all particles outside the cage. (d) Dimensionless mean alpha relaxation or hopping time (and its analog with no elastic barrier, dashed curve) for dense metastable hard sphere fluids as a function of packing fraction. The inset shows a cross-plot of the mean alpha time and elastic plateau shear modulus in the transiently localized state on a log-linear scale. (e) Exponential growth of the dimensionless elastic shear modulus (triangles) (units of $k_BT\sigma^{-3}$) associated with the transiently localized state, along with the maximum caging force (squares)  felt by a tagged particle. Inset shows a cross plot of $f_{max}$ versus $G^\prime$.
	} 
	\label{fig1}
\end{figure*}
We now briefly review all aspects of the transient dynamics and elasticity under quiescent conditions and their changes under deformation which serve as the starting point for our new theoretical advances. All technical details and physical discussions have been reported in prior papers  \cite{schweizer2005derivation-c99,kobelev2005strain-2be,mirigian2014elastically-4c6,ghosh2023microscopic-39e,mutneja2025microscopic-f31,mei2025mediumrange-05d}, including many successful confrontations with experiments  \cite{chaki2024theoretical-92e,athanasiou2025probing-0e7}.

The starting point for equilibrium dynamics is the force level ECNLE theory of stochastic activated single particle trajectories as encoded in a nonlinear Langevin equation (NLE): 
\begin{equation}\label{NLE}
-\zeta_s\frac{dr}{dt}-\frac{\partial F_{dyn}}{\partial r}+\delta f=0
\end{equation}
Eq.(\ref{NLE}) represents an overdamped (Brownian dynamics) stochastic force balance evolution equation for the angularly averaged scalar displacement $r(t)$ of a tagged particle. Here, $\delta f$ is the white noise random force that obeys $\left\langle\delta f\left(0\right)\delta f\left(t\right)\right\rangle=2k_BT\zeta_s\delta\left(t\right)$, and $\zeta_s$ is the short-time and distance frictional drag characterised by a well-known short-time friction constant  \cite{schweizer2005derivation-c99}. The associated time $\tau_s=\beta\zeta_s\sigma^2$ is the unit of time in all our calculations. It arises from the non-activated, in-cage, short-time and distance process that contains local hydrodynamic and solvent effects in colloidal suspensions and two-particle dynamics (collisions for hard spheres) in a mean field cage framework. The well-known explicit expression for hard spheres is provided in Appendix.\ref{ApdA1} (Eq.\ref{Taus}) and for details see Ref.\cite{schweizer2005derivation-c99}.

The key quantity in Eq.(\ref{NLE}) is the microscopic spatially-resolved dynamic free energy, $F_{dyn}\left(r\right)$ \cite{schweizer2005derivation-c99,mirigian2014elastically-4c6}, derived using a non-traditional form of dynamic density functional theory (DDFT). Its negative gradient determines the effective force on a particle due to all other moving particles: 
\begin{equation}\label{Fdyn}
	\begin{split}
\beta F_{dyn}\left(r\right)=&-3\ln{\frac{r}{\sigma}}-\\&\frac{\rho}{2\pi^2}\int_{0}^{\infty}{\frac{k^2C^2\left(k\right)S\left(k\right)}{1+S^{-1}\left(k\right)}e^{-\frac{k^2r^2}{6}\left(1+\frac{1}{S\left(k\right)}\right)}}dk
\end{split}
\end{equation}
Here $\sigma$ is the particle diameter, $\beta\equiv\left(k_BT\right)^{-1}$ the inverse thermal energy, $\rho$ is the number density, and $\phi\equiv\frac{\pi\rho\sigma^3}{6}$ is the dimensionless packing fraction. Dynamically relevant effective pair interactions (and hence forces) enter via the direct correlation function determined by equilibrium structure, which in Fourier space obeys the Ornstein-Zernike relation $C\left(k\right)=\rho^{-1}\left[1-S^{-1}\left(k\right)\right]$ where $S\left(k\right)$ is the static structure factor computable from accurate integral equation theory (see Appendix:\ref{ApdA2})\cite{zhou2020integral-c4b}. Physically, the dynamic free energy sums up slowly decaying force-force time correlations on all relevant length scales (wavevector integration in Eq.(\ref{Fdyn})) within a local equilibrium and non-ensemble-averaged formulation of DDFT  \cite{schweizer2005derivation-c99}. For repulsive particles, it defines the steric caging potential experienced by a tagged particle due to the motion of all surrounding particles (see Fig.\ref{fig1}(a)). However, the approach sketched above is general for any spherical particle system interacting via any pair potential.

Figure \ref{fig1}(b) shows an example quiescent dynamic free energy with all relevant length and energy scales indicated; a non-zero barrier emerges beyond a critical packing fraction $\phi_c=0.44$  \cite{zhou2020integral-c4b} for hard spheres. This localized form signals a dynamical crossover predicted from a simplified or naïve mode coupling theory (NMCT). The activated dynamics theory predicts the particle jump length $\Delta r$ required to cross the activation barrier of height $F_B$ and “escape its cage” is sufficiently large that a small collective elastic expansion of the cage is required, resulting in a spatially nonlocal contribution to the alpha relaxation event (see Fig.\ref{fig1}(a)). Such a spatially non-local extension defines the Elastically Collective NLE (ECNLE) theory  \cite{mirigian2014elastically-4c6} via the presence of a longer range elastic barrier ($F_{el}$) that is causally related to the local cage barrier as required to achieve the elementary irreversible barrier crossing event, as schematically shown in Fig.\ref{fig1}(c). Importantly, the dynamic free energy contains \textit{all} essential elements to compute $F_{el}$ (see Appendix:\ref{ApdA3} or  \cite{mirigian2014elastically-4c6} for detailed discussion), and the mean activated alpha relaxation time, $\tau_\alpha$. 

Given the particle interactions and the thermodynamic state, the dynamic free energy is a priori constructed. From it, the elastic shear modulus, $G^\prime$ (Eq.(\ref{GP}), microscopic Green-Kubo formalism) and $\tau_\alpha$ (Eq.(\ref{Tau}), Kramers mean first passage time) as follow as \cite{kramers1940brownian-150}. 

\begin{equation}\label{GP}
\frac{G^\prime}{\operatorname{k}_BT/\sigma^3}=\frac{k_BT}{60\pi^2}\int_{0}^{\infty}dk\left[k^2\rho S\left(k\right)\frac{dC(k)}{dk}\right]^2e^{-\frac{k^2r_L^2}{3S\left(k\right)}}
\end{equation}
\begin{equation}\label{Tau}
\frac{\tau_\alpha}{\tau_s}=e^{\beta F_{el}}\int_{r_L}^{r_B}{dx\ e^{\beta F_{dyn}(x)}}\int_{r_L}^{x}{dy\ e^{-\beta F_{dyn}(y)}}	
\end{equation}
Here, $r_L$ is the transient particle localization length (minimum of the dynamic free energy) and $r_B$ is the barrier location. Above and below we non-dimensionalize $G^\prime$ and stress, $\Sigma$, by the Brownian stress scale $\frac{k_BT}{\sigma^3}$, and $\tau_\alpha$ and ${\dot{\gamma}}^{-1}$ by $\tau_s$ discussed above \cite{schweizer2005derivation-c99}.

Fig. \ref{fig1}(d, e) shows the packing fraction variation for hard spheres of $\tau_\alpha$ and $G^\prime$ in the equilibrium highly metastable range of present interest. The inset of panel (d) demonstrates that ECNLE theory predicts the total activation barrier and dynamic elastic shear modulus are exponentially related as $ln{\left(\tau_\alpha/\tau_0\right)}\propto\beta\sigma^3G^\prime$. Other connections of the alpha time to the medium-range order correlation length and a specific density fluctuation measurable thermodynamic property ($S(k=0)\equiv S_0$) have been derived, and shown to be in accord with experiments on thermal glass forming molecular and polymer liquids  \cite{mei2025mediumrange-05d}. Panel (e) also shows the packing fraction variation of the maximum caging force, $f_{max}$, felt by a particle that occurs at the inflection point displacement of the highly anharmonic dynamic free energy; to leading order in the deeply metastable regime for hard spheres  \cite{schweizer2007collisions-33c}, $f_{max}\propto\frac{k_BT}{r_L}$.  The inset shows a cross plot of this quantity against the elastic modulus, which follows from the microrheology-like prediction that $G^\prime\propto k_BT/\sigma r_L^2$. Taken as a whole, these plots illustrate the deep connection among all properties of the spatially resolved microscopic dynamic free energy via their common origin in spatially-resolved structure-determined kinetic constraints, from which ECNLE theory predicts dynamic observables, as discussed in great detail  \cite{schweizer2005derivation-c99,mirigian2014elastically-4c6}. 

The above description of glass-forming liquids \textit{in equilibrium} is well-established and has been recently generalized to self-consistently treat non-linear rheology, and successfully utilized to predict the nonlinear mechanical response to external deformation  \cite{ghosh2023microscopic-39e,mutneja2025microscopic-f31}  as well as re-equilibration dynamics after deformation cessation \cite{AMJOR2026}. The latter are achieved in a tractable and predictive manner via the incorporation of two emergent deformation variables: the macroscopic stress, $\Sigma$, and the microscopic structural deformation as will be quantified by the effective scalar strain, $\gamma_{eff}$. Physically, the stress accounts for the particle-level unbalanced forces, while the effective strain accounts for the change in microscopic structure and its influence on dynamics and rheology. These two quantities are zero in the quiescent state at the start of the deformation, and self-consistently evolve over time as we now discuss. 

\begin{figure}
	\includegraphics[width=0.45\textwidth]{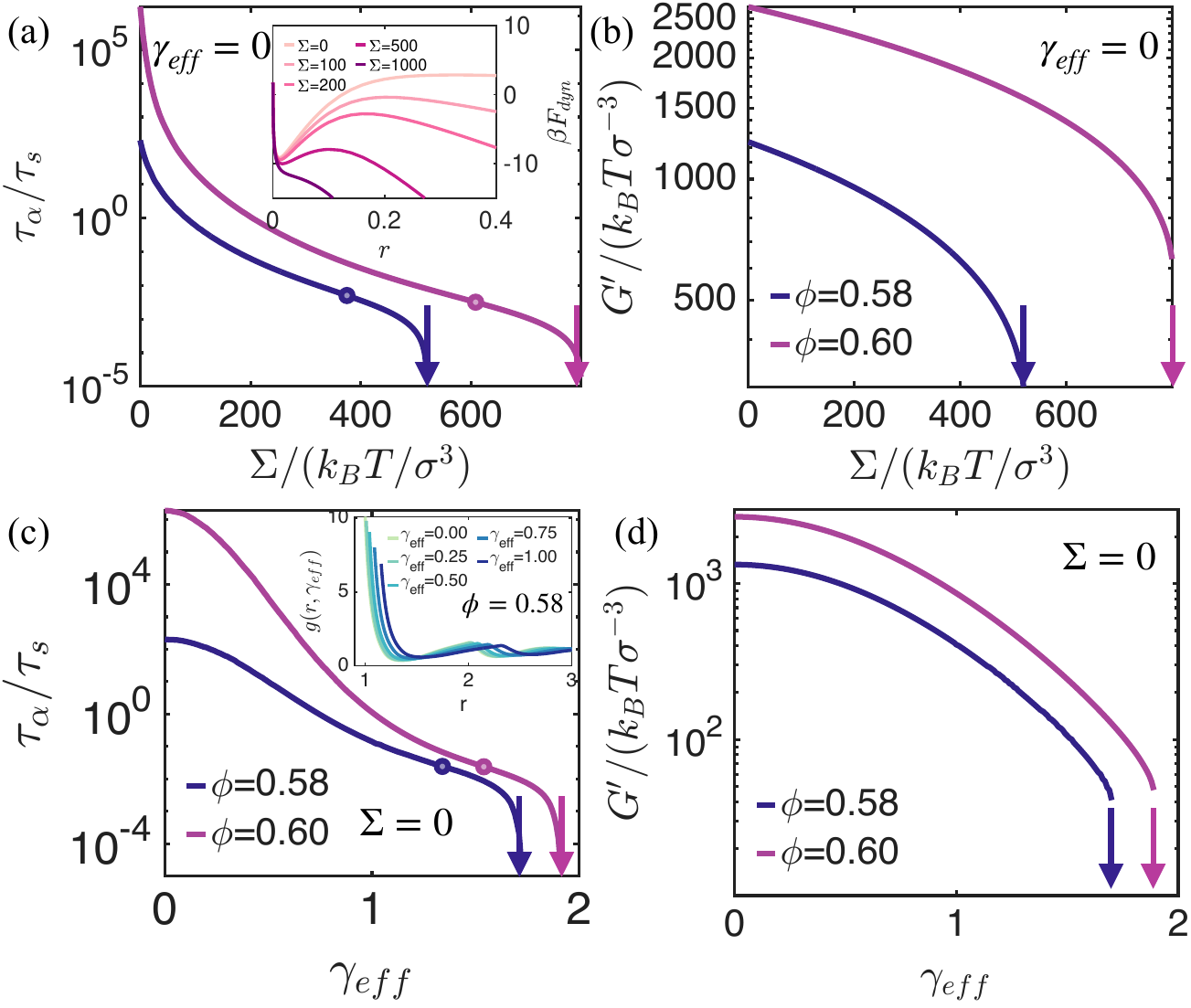}
	\caption{(a) Parametric evolution of $\tau_\alpha$ with stress in the idealized isostructural $\gamma_{eff}=0$ limit. Circles mark the stress where the barrier is reduced to $k_BT$, while arrows are when it strictly vanishes defining the ``absolute yield stress” limit. The inset shows the dynamic free energy as a function of increasing stress. (b) Corresponding elastic modulus evolution. (c) Parametric evolution of $\tau_\alpha$ with effective affine strain in the idealized limit of zero stress. The inset shows the corresponding change of the pair correlation function, $g(r)$ based on wavevector advection. Arrows mark the absolute yield strain where the localized state is destroyed. (d) Corresponding elastic modulus evolution.
	} 
	\label{fig2}
\end{figure}

The macroscopic stress enters the theory based on the microrheological idea that it is transduced to the microscopic particle level as a mechanical force, yielding a nonequilibrium work-type contribution to dynamic free energy  \cite{kobelev2005strain-2be}: $F_{dyn}\left(r;\Sigma\right)=F_{dyn}\left(r;\Sigma=0\right)-\frac{\pi\sigma^2}{24}\Sigma\ r$, where the microscopic prefactor is proportional to a single particle cross-sectional area. Because of the anharmonic and finite barrier form of the dynamic free energy, the effective mechanical force or stress monotonically reduces all measures of solidity (Fig.\ref{fig2}(a, b)), including the transient localization (Fig.\ref{fig2}(a)-inset), elastic modulus, and activation barrier \cite{kobelev2005strain-2be}.

The internal structural deformation is microscopically quantified via the scalar effective strain variable $\gamma_{eff}$ based on the \textit{no} adjustable parameter \textit{isotropic} wavevector advection idea, $S\left(k,\gamma_{eff}\right)\rightarrow S\left(k\sqrt{1+\frac{\gamma_{eff}^2}{3}}\right)$ \cite{amann2014transient-b79,fuchs2009high-399}. This is a second origin of kinetic constraint ``softening” encoded in the dynamic free energy, Fig.\ref{fig2}(c, d). The inset of Fig.\ref{fig2}(c) shows the evolution of the pair correlation function, $g\left(r\right)$, with increasing $\gamma_{eff}$. The suppression in amplitude and increase in interparticle separation of its near‑contact region reflect deformation-induced cage stretching. 

The rationale for the adopted effectively isotropic treatment of structural deformation has been discussed in great depth previously  \cite{ghosh2023microscopic-39e,mutneja2025microscopic-f31}. It is adopted for technical tractability since a reliable microscopic theoretical description of the fully anisotropic structure factor is not available. Even if it was, the associated tensorial description of kinetic constraints, the dynamic free energy, and the rheological response would be computationally and conceptually much more complex. In addition, a physical motivation is that both simulations  \cite{koumakis2012yielding-592}  and experiments  \cite{besseling2007threedimensional-9cf} provide support for the adequacy of this simplification on the \textit{local} length scales most relevant to the theoretical predictions based on the dynamic free energy concept. From a broader tensorial viewpoint, the resulting elastic and viscous softening arises mainly from extensional‑axis–dominated distortions, an averaged isotropic effect of which is approximately captured by the wave-vector advection (Fig.\ref{fig2}(c)-inset). At very high strain rates, however, strong compression‑axis densification may induce nontrivial effects, such as shear jamming \cite{pan2023review-d54}, which is beyond the reach of the present isotropic framework.

Under nonequilibrium conditions, the analogs of Eqs. (\ref{GP}) and (\ref{Tau}) for the elastic modulus and relaxation time continue to hold in form, except the required structural input is a function of $\gamma_{eff}$, and the localization length and dynamic free energy depend on both stress and effective strain,  $F_{dyn}(r;\Sigma,\gamma_{eff})$. Thus, microscopic ECNLE theory predictions of the alpha time and elastic modulus become highly nonlinear two-dimensional parametric functions of stress and effective strain. For pedagogical reasons, and also to clearly expose the new physics under deformation in the theory, we first present example calculations for metastable hard sphere fluids that illustrate the isolated parametric influence of each deformation variable discussed above.

Figure \ref{fig2} (a, b), respectively, shows the \textit{parametric} influence of external stress on the relaxation time and elastic modulus under the idealized condition of fixed structure ($\gamma_{eff}=0$). External stress enters directly as an extra force at the particle level, which progressively weakens \textit{all} localizing features of the anharmonic dynamic free energy. With increasing stress, the activation barriers are initially reduced on the larger length scale that defines the origin of the collective elastic barrier, an effect that is amplified as packing fraction increases. With further stress increase, the local cage barrier is also reduced, ultimately leading to a drastic reduction in the relaxation time. Once the \textit{total} barrier approaches the thermal energy scale at a high enough stress (circles in Fig.\ref{fig2}(a)), a sharp drop of the relaxation time occurs. This marks a mechanically-induced delocalization transition (shown by arrows) at a characteristic stress called the ``absolute yield stress” where the dynamic free energy no longer displays any localizing features. In contrast, since the elastic modulus is governed by the much shorter localization length scale of the dynamic free energy, the softening effects due to external stress are far more modest and evolve in a smoother manner.

Figure \ref{fig2} (c, d) shows the kinetic constraint softening effects of strain-induced changes of structure on the relaxation time and elastic modulus under the idealized condition of zero applied stress. Similar to the isolated effect of external stress, reduction of the relaxation time is more pronounced than elastic modulus softening, and the corresponding ``absolute yield strain” (arrows) again marks the delocalization transition. However, compared with the effect of external stress, the overall softening induced by structural deformation is weaker at small deformations. This difference arises because structural deformation modifies the dynamic free energy through an integral over time-persistent force correlations over all length scales in Eq.(\ref{Fdyn}), while external stress enters the dynamic free energy directly as a particle-level mechanical work term. Of course, in a real step-shear deformation the effects of stress and strain are simultaneously present in a correlated manner.

We now build on all of the above to construct a coupled theory for the mechanical constitutive equation and nonequilibrium structural evolution equation. The system response, as in the coupled stress and structural evolution based on a time-local system, is obtained by self-consistently solving the following generalized nonlinear Maxwell model for homogeneous deformations of \textit{coupled} stress and structural evolution  \cite{ghosh2023microscopic-39e,mutneja2025microscopic-f31,chen2011theory-f68}. 

\begin{equation}\label{Maxwell}
\frac{d\Sigma\left(t\right)}{dt}+\frac{\Sigma\left(t\right)}{\tau_\alpha\left[\Sigma\left(t\right),\gamma_{eff}\left(t\right)\right]}=\ \dot{\gamma}G^\prime\left[\Sigma\left(t\right),\gamma_{eff}\left(t\right)\right]
\end{equation}
\begin{equation}\label{GammaEff}
	\frac{d\gamma_{eff}\left(t\right)}{dt}=-\frac{\gamma_{eff}\left(t\right)}{\tau_\alpha\left[\Sigma\left(t\right),\gamma_{eff}\left(t\right)\right]}+\dot{\gamma}
\end{equation}
$\Sigma\left(t\right)$ is the stress acquired elastically via the dynamic shear elastic modulus, $G^\prime\left(t\right)$, which is dissipated on a nonequilibrium alpha time scale, $\tau_\alpha\left(t\right)$, with the accumulated shear strain $\gamma\left(t\right)\equiv\dot{\gamma}t$ acquired at a constant shear rate $\dot{\gamma}$. Eq. (\ref{GammaEff}) describes the evolution of the microscopic structural deformation parameter $\gamma_{eff}$, which increases elastically via the affine total strain (last term) and competes with a return to the quiescent state via activated relaxation on a timescale $\tau_\alpha(t)$ (first term). Importantly, $\gamma_{eff}$ is not the trivial affine strain per a purely elastic ideal solid, but includes activated irreversible relaxation (plastic effects), and at any time during the deformation quantifies the non-quiescent structure and thus serves as a microstructural deformation memory. 

Here we propose $\gamma_{rec}\equiv\gamma_{eff}$ describes the recoverable strain, which is of central conceptual importance in recovery rheology. Its difference from the total macroscopic strain defines the plastic unrecoverable strain, $\gamma_{unrec}\equiv\gamma-\gamma_{rec}$. 

We emphasize that Eqs. (\ref{Maxwell}) and (\ref{GammaEff}) have implicitly assumed that the relevant stress and structural relaxation times are exactly equal. This is the minimalist description, and for activated glassy dynamics in equilibrium, it is typically true to within a prefactor that deviates little from unity  \cite{denisov2013resolving-853}. However, this is a nontrivial, largely unsolved problem the precise quantification of which depends on the specific system. The modelling of this nonuniversal physics via a so-called constant (independent of thermodynamic state and deformation variables) ``mismatch factor", $f=\tau_\alpha^{Stress}/\tau_\alpha^{Structure}$, can be introduced, as in Appendix \ref{ApdB}.

The coupled self-consistent Eqs (\ref{Maxwell}) and (\ref{GammaEff}) are combined with the ECNLE theory computed $G^\prime\left[\Sigma\left(t\right),\gamma_{eff}\left(t\right)\right]$ and $\tau_\alpha\left[\Sigma\left(t\right),\gamma_{eff}\left(t\right)\right]$ to provide a closed description that is numerically solved to obtain the nonequilibrium stress and structure response. We emphasize that the multiple integrated theoretical elements have been extensively tested against experiment \cite{ghosh2023microscopic-39e,mutneja2025microscopic-f31} \textit{without fitting} for the step-rate start-up shear response of dense glass-forming hard particle fluids. We now apply the theory to provide new structural, dynamical, and rheological predictions, and most importantly, obtain new inter-relations relevant to recovery rheology for metastable hard sphere suspensions. Very importantly, these inter-relations are \textit{broadly applicable beyond hard spheres}. As proof of concept, analogous findings for soft particles modelled by a repulsive Hertzian potential   \cite{yang2011glassy-b3f} are presented in Section \ref{Sec4}.

\section{Results}
We begin by presenting foundational theoretical results, followed by new predictions for the connections between recoverable and unrecoverable strain with steady state shear thinning, the transient stress overshoot, and the structural and stress relaxation time. This includes testable inter-relationships between rheological features on different timescales or equivalently total accumulated strain predicted by our approach. 
\subsection{Foundational Behaviors}\label{Sec3a}
\begin{figure}[t]
	\includegraphics[width=0.45\textwidth]{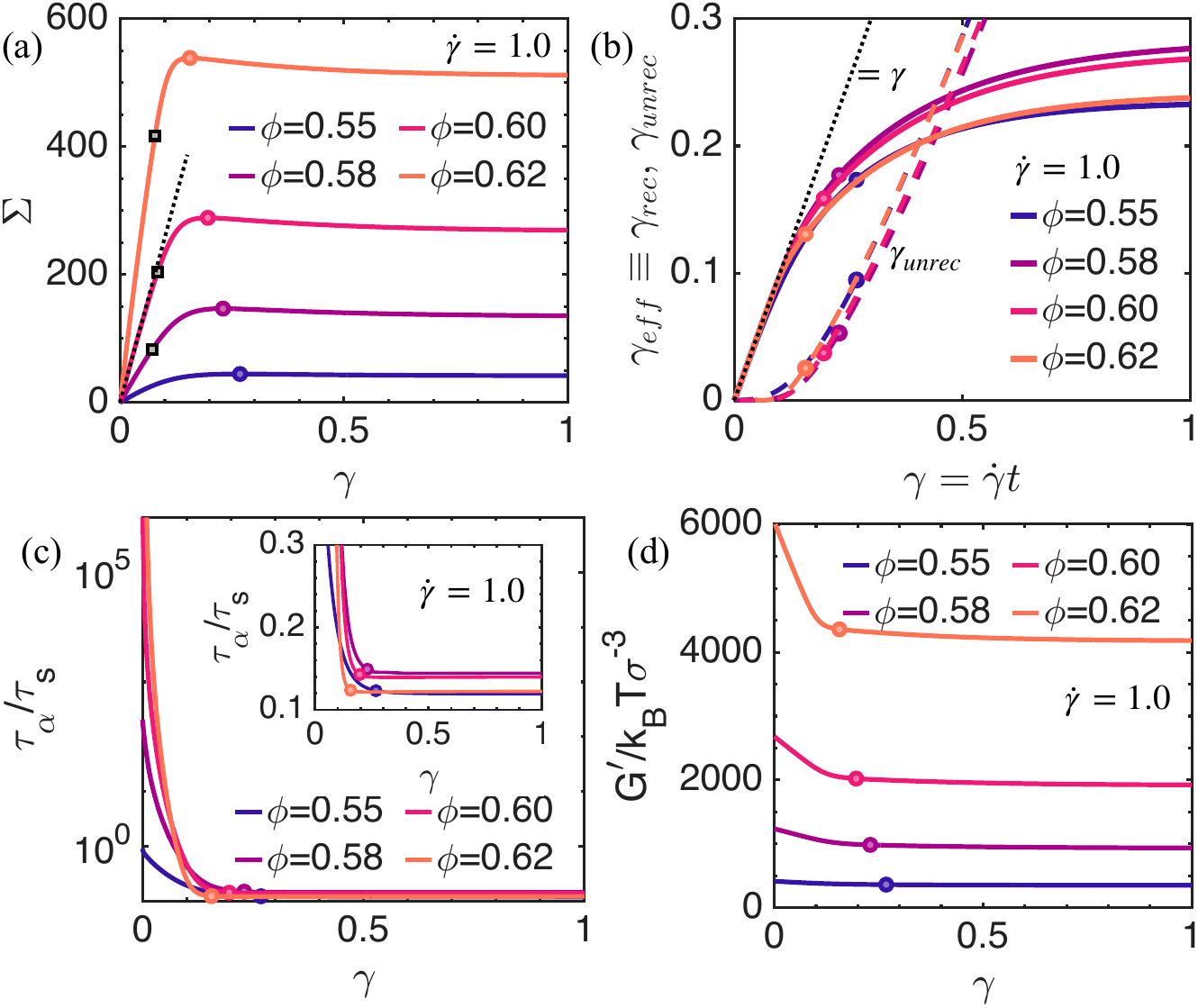}
	\caption{(a)  Stress-strain response for four packing fractions and a dimensionless shear rate of $\dot{\gamma}=1$. Points mark the stress overshoot. The renormalized Peclet number defined as $Pe_0\equiv\tau_\alpha\left(0\right)\dot{\gamma}$ for the quiescent alpha times of $0.9,\ 208,\ 2\times{10}^6,\ 4\times{10}^{16}$ corresponding to $\phi=0.55,\ 0.58,\ 0.60,\ 0.62$ decreases to unity at the indicated squares marking the onset of plasticity ($Pe\left(\gamma\right)\equiv\tau_\alpha\left(\gamma\right)\dot{\gamma}=1$), where the strain is of order $0.1$. (b) Corresponding structural strain parameter evolution (recoverable strain, solid lines), which at the overshoot remains far below its steady state value, and unrecoverable strain (dashed lines). The corresponding dimensionless alpha time (c) and elastic modulus evolution (d) as a function of strain. }
	\label{fig3} 
\end{figure}
Figure \ref{fig3} shows the concurrent evolution of the stress, the structural recoverable strain, the alpha relaxation time, and the elastic plateau modulus for four metastable high packing fractions. Initially, when the alpha time exceeds the inverse shear rate per a solid-like system (renormalized Peclet number $Pe_0\equiv\tau_\alpha\left(t=0\right)\dot{\gamma}\approx{10}^6\dot{\gamma}$ for $\phi=0.6$), the deformation-modified activated relaxation effects on the rheological response are minimal, \textit{despite} the fact that modest deformation \textit{does} strongly reduce the relaxation time (but $Pe(t)\equiv\dot{\gamma}\tau_\alpha(t) >>1$). As a result, $\Sigma$ and $\gamma_{eff}$ increase linearly with time at their respective rates of $\dot{\gamma}G^\prime(t)$ and $\dot{\gamma}$, marking the operationally defined elastic regime. Further increase in deformation eventually sufficiently weaken the caging constraints, strongly reducing the elastic modulus and relaxation times and rendering activated processes relevant on the inverse shear-rate timescale. This results in significant plastic effects and an anelastic regime; e.g., $Pe\left(t\right)\approx1$ ($0.14$) at plastic onset (overshoot) shown by square (circle) for $\phi=0.6$ sheared at rate $\dot{\gamma}=1.0$ in Fig.\ref{fig3}(a). The stress then undergoes a non-monotonic overshoot before reaching a steady state. This behavior is predicted to be in striking qualitative contrast to the structural strain behavior in Fig.\ref{fig3}(b), which more slowly and monotonically approaches the steady state. In the long-time steady state, viscous stress relaxation matches elastic stress generation, and structural relaxation matches structural deformation, while the full transient behavior depends on the instantaneous values of these four contributions. 

We now decompose the macroscopic total strain into its recoverable (${\gamma_{rec}=\gamma}_{eff}$) and unrecoverable $\gamma_{unrec}=\gamma-\gamma_{eff}$ components. The latter corresponds to particle displacements accumulated opposite to the shear direction and thus quantifies the plastic  response. As shown in Fig.\ref{fig3}(b) (dashed lines), it starts from zero and remains small until the overshoot, after which it increases with strain rate. Another important facet that emerges from recognizing the effective strain as the recoverable strain is the resolution of  the applied strain rate into its  recoverable and unrecoverable components. At early times, when relaxation processes are negligible, the microstructure responds affinely so that ${\dot{\gamma}}_{eff}(t\rightarrow0)\equiv{\dot{\gamma}}_{rec}(t\rightarrow0)\approx\dot{\gamma}$. In contrast, in the steady state, where the microstructure  is out of equilibrium but has ceased to evolve, the deformation is fully plastic, and thus ${\dot{\gamma}}_{unrec}(t\rightarrow\infty)\approx\dot{\gamma}$. The transient evolution, however, will be shown in Sec.\ref{Sec3e} to be linked and in a manner that allows one to obtain transient alpha time evolution. 

Figure \ref{fig3}(c) shows that the relaxation time is predicted to decrease quickly from $\left[0.9,\ 208,\ 2\times{10}^6,\ 4\times{10}^{16}\right]$ for packing fractions of $\left[0.55,\ 0.58,\ 0.60,\ 0.62\right]$, and becomes almost independent of time/strain around the overshoot (see inset for zoomed-in plot). In contrast, the stress and structure continue to evolve. This clearly establishes that the overshoot feature marks a yielding or solid-fluid transition from a structural or stress relaxation time perspective. Similarly, the elastic modulus in Fig. \ref{fig3}(d) displays a significant reduction until overshoot, and then slowly saturates to a steady-state value. 

\subsection{Connecting Shear-Thinning Response and Recoverable Strain}\label{Sec3b}
\begin{figure*}[t]
	\includegraphics[width=0.95\textwidth]{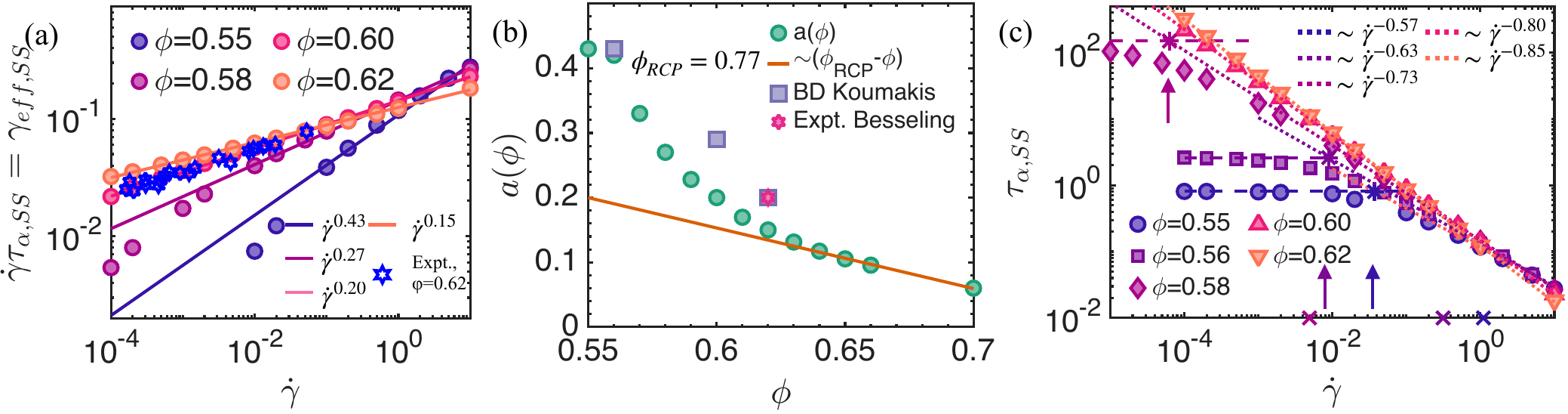}
	\caption{(a) Predicted steady state shear thinning response for different packing fractions in the format of the steady state Peclet number $\dot{\gamma}\tau_{\alpha,SS}=\gamma_{rec,SS}\sim{\dot{\gamma}}^{a\left(\phi\right)}$; apparent power law scalings are indicated. The stars indicate experimental data from Ref. \cite{besseling2007threedimensional-9cf}  which follow an effective power law of ${\dot{\gamma}}^{0.2}$ for polymethylmethacrylate hard colloids with $10\%$ polydispersity and $\phi\sim0.62$. (b) Packing fraction dependence of exponent $a\left(\phi\right)$, which linearly approaches zero as the approximate theoretical random close packing (RCP) state is approached at $\phi_{RCP}$ \cite{chaki2024theoretical-92e} (see Appendix:\ref{ApdA2}) following a critical scaling form of $a\sim\left(\phi_{RCP}-\phi\right)^1$. The squares are the exponents extracted from Ref.\cite{koumakis2011study-b2a}  from the Brownian dynamics simulations, and the star is the experimental exponent from Ref. \cite{besseling2007threedimensional-9cf}.  (c) Dimensionless steady state alpha time as a function of shear rate $\dot{\gamma}$, showing effective power law behaviour as dotted lines in the large shear rate limit, and a Newtonian plateau (dashed lines) for small  $\dot{\gamma}$. The arrows and stars mark the onset of shear-thinning, while the crosses on the x-axis mark the inverse quiescent alpha times. See text for details.}
	\label{fig4} 
\end{figure*}
Based on the above results, our first new insight is the deduction of a quantitatively precise connection between recoverable strain and the alpha (or stress) relaxation time in the shear-thinned steady state. To see this, first note that in the steady state Eq.(\ref{GammaEff}) simplifies to $\gamma_{rec, SS}\equiv\gamma_{eff,SS}=\dot{\gamma}\tau_{\alpha,SS}$, where $\tau_{\alpha,SS}$ is the steady state structural relaxation time directly related to the steady state recoverable strain scaled by ${\dot{\gamma}}^{-1}$. This intriguing connection between the viscous shear-thinning response and the recoverable strain has been deduced microscopically and in a \textit{general} manner from the structure of our theory in Eqs. (\ref{Maxwell}) and (\ref{GammaEff}), and is testable by macroscopic rheology experiments. The generality of this expression stems from a simple physical idea: in the steady state, the structural relaxation rate, $\frac{\gamma_{rec,SS}}{\tau_{\alpha,SS}}$, must balance the applied deformation rate $\dot{\gamma}$ to prevent further microstructural evolution. However, the detailed mechanism of structural deformation, such as wave‑vector advection, is secondary. 

The above finding also provides an understanding of the origin of deviations from the ideal shear-thinning response of $\tau_{\alpha,SS}\sim{\dot{\gamma}}^{-1}$ as arising from the shear rate dependence of the structural cage deformation. More specifically, from Fig. ref{fig4}(a) one sees that the recoverable strain is predicted to increase as a power law of $\gamma_{rec}\sim{\dot{\gamma}}^{a\left(\phi\right)}$ with an exponent $a\left(\phi\right)\in\left[0-1\right]$. This leads to the steady-state alpha-time (and to leading order the non-Newtonian viscosity) obeying $\tau_{\alpha,SS}\sim{\dot{\gamma}}^{-\left(1-a\left(\phi\right)\right)}$, with $a(\phi$) systematically approaching zero with increasing packing fraction. The packing fraction dependence of the extracted theoretical exponent is shown in Fig.\ref{fig4}(b) to linearly approach zero as random close packing (RCP) of the employed integral equation theory is approached: $a(\phi\rightarrow\phi_{RCP})\sim\left(\phi_{RCP}-\phi\right)^1$; see Appendix:\ref{ApdA2} or Ref. \cite{chaki2024theoretical-92e}  for detailed discussion on the RCP limit. Physically, it implies that, as $\phi\rightarrow\phi_{RCP}$ (per frictionless granular systems), the cage deformation required for flow becomes strain rate \textit{independent}, per our prediction of ideal shear thinning. Conversely, at lower more weakly metastable packing fractions, steady-state cage deformation is further enhanced with increasing shear rate, implying a smaller reduction in the alpha time is needed to attain steady-flow and thus a smaller thinning exponent. 

Importantly, an effective shear thinning exponent smaller than unity has been experimentally observed in steady shear  \cite{besseling2007threedimensional-9cf}  (compared in Fig.\ref{fig4}(a) and the single point in Fig.\ref{fig4}(b)), along with smaller fractional powers with decreasing $\phi$ for particle self-diffusion coefficients during large amplitude oscillatory shear  \cite{koumakis2011study-b2a,koumakis2013complex-b0e} (squares in Fig. \ref{fig4}(b)); see figure captions for system details. The trends and absolute values from simulation and theory agree well, especially considering the theoretical calculations based on smooth hard spheres do not include polydispersity effects and other real-world complications such as colloid surface roughness. Interestingly, similar responses have also been seen in polymeric glasses with increasing temperature \cite{hebert2015effect-983}  and in numerical simulations of model supercooled/glassy systems \cite{berthier2002nonequilibrium-005,mizuno2024universal-6d9}. Moreover, numerous studies  \cite{eisenmann2009shear-e56,jacob2015convective-227} on extremely dense or soft jammed systems have shown near ideal shear-thinning behavior. Our theory provides a microscopic physical basis for these behaviors based on the strain-rate variation of the steady-state recoverable strain that underlies apparent scaling exponent variability. 

Figure \ref{fig4} (c) presents the predicted evolution of the steady state alpha relaxation time from its Newtonian plateau quiescent value at low shear rates  to an inverse‑effective fractional power‑law regime characteristic of traditional shear thinning. For relatively low, but still metastable fluid, packing  fractions, the Newtonian plateau at small shear rates smoothly crosses over to a power‑law thinning regime where $\tau_{\alpha,SS}\sim{\dot{\gamma}}^{-(1-a\left(\phi\right))}$ at shear rates indicated by the arrows. Strikingly, the crossover shear rate is decades smaller than the inverse quiescent alpha  relaxation time, shown by the crosses on the x‑axis. Specifically, for $\phi=[0.55,\ 0.56,\ 0.58]$, the corresponding renormalized Peclet numbers are $Pe=[0.0337,\ 0.0285,\ 0.0128]$, approximately two decades below unity, and they decrease further for more deeply metastable states. This behavior is  consistent with experimental and simulation studies \cite{mizuno2024universal-6d9,webb1990onset-188}, and very different from the shear thinning behavior of entangled polymer melts  \cite{rubinstein2003polymer-ca0,doinoyeartheory-bb4}  but akin to that of polymer glasses  
\cite{chen2008microscopic-ff1,xing2022segmental-4a3}. As discussed earlier, the effective large‑strain‑rate shear‑thinning exponents approach the ideal value of $-1$ as the packing fraction increases.

 \subsection{Connecting Transient Overshoot Magnitude and Steady-state Recoverable Strain}\label{Sec3c}
 A second new insight from our structure-based theoretical recovery rheology viewpoint follows from considering the transient overshoot under the homogeneous deformation conditions of present interest (no spatially inhomogeneous shear-banding). To explain our findings in the simplest way, first 
 consider a situation where one assumes $G^\prime$ does \textit{not} soften under deformation (per the KDR model \cite{kamani2021unification-578}). It is easy to show that \textit{no} overshoot is predicted, and stress and structure follow a simple monotonic plastic-like response. This occurs since Eqs. (\ref{Maxwell}) 
 and (\ref{GammaEff}) become slaved with a multiplicative factor of the constant elastic modulus, yielding $\Sigma\left(t\right)=\gamma_{eff}\left(t\right)G^\prime$. Thus, if $\frac{d\Sigma}{dt}=0$, then $\frac{d\gamma_{eff}}{dt}=0$, leading to 
 $\frac{d^2\Sigma}{dt^2}=\frac{1}{\tau_\alpha^2}\frac{d\tau_\alpha}{dt}=0$ since $\tau_\alpha\left(t\right)=\tau_\alpha(\Sigma\left(t\right),\gamma_{eff}(t))$. On the other hand, in reality, our microscopic theory predicts deformation-induced $G^\prime$ softening (see Fig. \ref{fig3}(d)). This reduces the 
 stress generation rate over time, resulting in reaching the stress relaxation rate \textit{earlier} (at overshoot) than when the structural relaxation rate matches the deformation rate (at steady state). Thus, mechanistically, the structure continues to deform (see Fig. \ref{fig3}(b)), leading to more stress 
 relaxation and a stress overshoot. In qualitative physical terms, because nonlinear elastic softening leads to slower stress generation, structural deformation as externally driven will always extend in time and strain beyond the maximum stress storing capability of the material, leading to stress reduction in 
 time and the overshoot. The overshoot magnitude will thus be larger for a more deformed structural cage (see below). 
 
 The above mechanism for the stress overshoot is predicted to be dramatically enhanced if structural relaxation is quantitatively slower than its stress analog ($f<1$), leading to a larger steady-state deformation and more brittle behavior as defined as a larger overshoot (see Appendix. \ref{ApdB}, Fig. \ref{fig9}). Such behavior has been observed in dense repulsive soft colloid suspensions composed of crosslinked microgels  \cite{koumakis2012direct-9bc}  illustrating the generality of present theory beyond hard spheres. For example, ref.  \cite{koumakis2012direct-9bc}  reports a twofold increase in the stress overshoot amplitude for soft particles compared to the analogous hard colloid system. Such a value is comparable to that shown in Fig. \ref{fig9} of Appendix \ref{ApdB}  for $f=0.5$ corresponding to an elementary structural relaxation rate only one half that of the stress relaxation rate in Eqs (\ref{Maxwell}) and (\ref{GammaEff}), and also for the soft colloidal system in Fig. \ref{fig8} discussed in Sec.\ref{Sec4}. 
 \begin{figure}[t]
 	\includegraphics[width=0.4\textwidth]{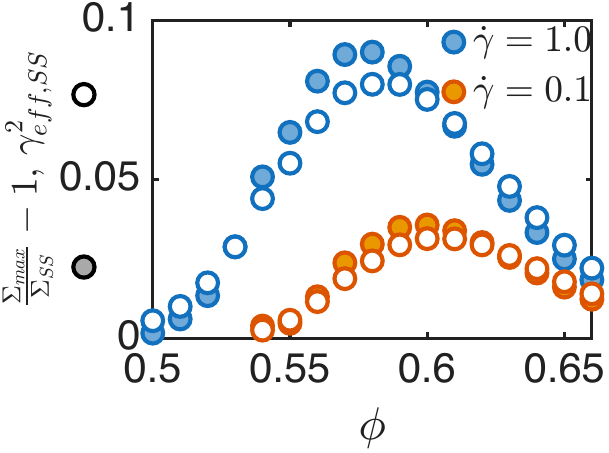}
 	\caption{ Filled symbols show the stress overshoot magnitude as a function of $\phi$ for two different $\dot{\gamma}$, along with its comparison to the effective \textit{steady state} structural strain squared $\gamma_{eff,SS}^2$ (open symbols). A remarkable near overlap is predicted, linking the nonlinear physics of the transient overshoot regime to that of the long-time nonequilibrium steady state.
 	} 
 	\label{fig5}
 \end{figure}
 
 A key and novel finding of Figure \ref{fig5} is that our microscopic theory quantitatively predicts a direct correlation between the transient overshoot magnitude, $\Sigma_{Max}/\Sigma_{SS}-1$ and the steady state structural deformation parameter squared, $\gamma_{rec,SS}^2\equiv\gamma_{eff,SS}^2=\left(\dot{\gamma}\tau_{\alpha,SS}\right)^2\equiv{{Pe}_{SS}}^2$. Interestingly, the overshoot follows a \textit{non-monotonic} trend with increasing packing fraction, which can be clearly understood from the predicted different behaviors in the low and high $\phi$ limits. At lower packing fractions approaching from above the ideal (naïve) MCT glass transition at $\phi=0.44$ that defines the onset of transient localization, emergent rigidity, and activated motion in ECNLE theory, the system fluidizes at very small external deformation since the cage restoring force of the dynamic free energy is predicted to approach zero \textit{continuously}  \cite{kobelev2005strain-2be}. Thus, in this limit, the system will yield plastically with a minimal overshoot and net structural deformation. In the opposite limit of high $\phi\rightarrow\phi_{RCP}$, the available open space for particle reconfiguration, and consequently the net allowed structural deformation, is sterically strongly reduced. Thus, the localization length (maximum cage restoring force of the dynamic free energy) is predicted to approach zero (infinity), while the elastic modulus for hard spheres approaches infinity since $G^\prime\propto k_BT\sigma^{-1}r_L^{-2}$, in the quiescent state  \cite{kobelev2005strain-2be}.  Hence, the yield strain (per a ratio of stress to elastic modulus) and overshoot amplitude will approach zero. Importantly, the linearly decreasing overshoot magnitude in the high $\phi$ regime has been seen experimentally, as previously suggested theoretically  \cite{ghosh2023microscopic-39e}. However, the striking non-monotonic trend observed experimentally  \cite{koumakis2012yielding-592}, along with its microscopic correlation with steady-state structure or renormalized Peclet number, is theoretically established here for the first time. This was qualitatively speculated previously based on simulation  \cite{marenne2017unsteady-1ac,kulkarni2009ordering-fe0,koumakis2012yielding-592}, and our derived precise connection to the Peclet number now renders it a testable prediction.
 
 The generality of the above correlation beyond the hard sphere system is further supported by both its theoretical  validation in soft‑particle systems as discussed in section \ref{Sec4} [Fig.\ref{fig8}(b)], and in hard sphere suspensions with slower elementary structural relaxation relative to stress relaxation (Appendix \ref{ApdB}, Fig.\ref{fig9}(c)).  Both of these systems exhibit overshoot magnitudes nearly twice those of the baseline hard-sphere model. 
\subsection{Unrecoverable Strain as a Universal Measure of Plasticity}\label{Sec3d}
\begin{figure}[t]
	\includegraphics[width=0.4\textwidth]{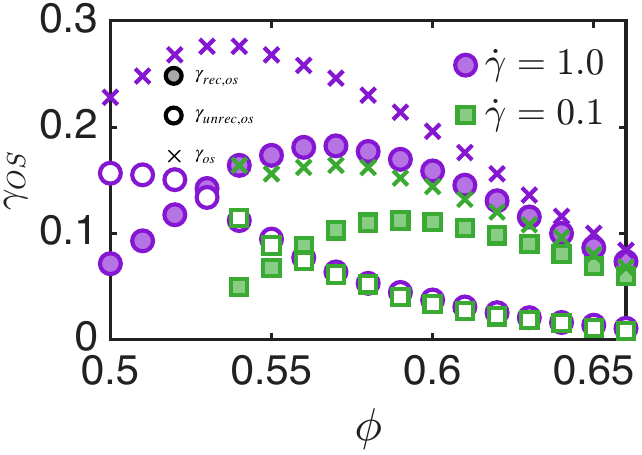}
	\caption{ Packing fraction dependence of the overshoot strains (crosses) for a step-rate deformation at two different strain rates, and with its separation into recoverable (solid) and unrecoverable (open) contributions. The total and recoverable overshoot strains exhibit non-monotonic packing fraction dependences, for the same physical reason discussed for the stress overshoot magnitude. In contrast, the unrecoverable strain monotonically decreases and is remarkably strain-rate independent. This leads to the conclusion that the overshoot (or yield point signalling a continuous solid-to-fluid transition) is associated with the material attaining a particular value of unrecoverable strain.  
	} 
	\label{fig6}
\end{figure}
The growth of unrecoverable strain with time shown in Fig.\ref{fig3}(b) in conjunction with our results for the overshoot strain value yields additional insights. Specifically, Fig.\ref{fig6} (and Fig.\ref{fig8}(c) for soft particles) shows the packing fraction variation of the overshoot strain and its separation into recoverable and unrecoverable components. The total  \cite{koumakis2008effects-0e9}  and recoverable \cite{petekidis2002yielding-8a3}  overshoot strain display a non-monotonic $\phi$ dependence for the \textit{same physical reasons} as the overshoot stress magnitude, while in contrast, the unrecoverable strain at the overshoot monotonically decreases. Perhaps even more intriguingly and unexpectedly, the unrecoverable strain is shear-rate \textit{independent}, rendering it an \textit{intrinsic} material property. Thus, the non‑monotonic packing‑fraction dependence of the overshoot strain arises from the corresponding non‑monotonic variation of the recoverable strain (i.e., the microscopic cage deformation at yield point), whereas the plastic strain required to yield (quantified by $\gamma_{unrec}$) decreases monotonically with $\phi$. We can thus conclude that the overshoot yield point feature in a step-rate experiment reflects the system acquiring a characteristic, volume-fraction-dependent critical unrecoverable strain. This serves as a new Lindemann-like criterion for yielding in terms of unrecoverable strain. 

Finally, the above discussion and Fig.\ref{fig6} also provide deeper physical insight into the strain-rate-dependent mechanistic definition of the emergence of solidity, which is relevant to experiments and the non-monotonic dependence of the yield strain. For lower packing fractions, where the overshoot strain grows with packing fraction, the unrecoverable strain is larger than its recoverable analog, corresponding to a more fluid-like response. In qualitative contrast, for packing fractions where the overshoot strain decreases with increasing $\phi$, then $\gamma_{unrec}<\gamma_{rec}$ per a more solid-like response. 

  \subsection{Predicting the Transient Alpha Time from the Unrecoverable Strain-Rate}\label{Sec3e}
  \begin{figure}[t]
  	\includegraphics[width=0.48\textwidth]{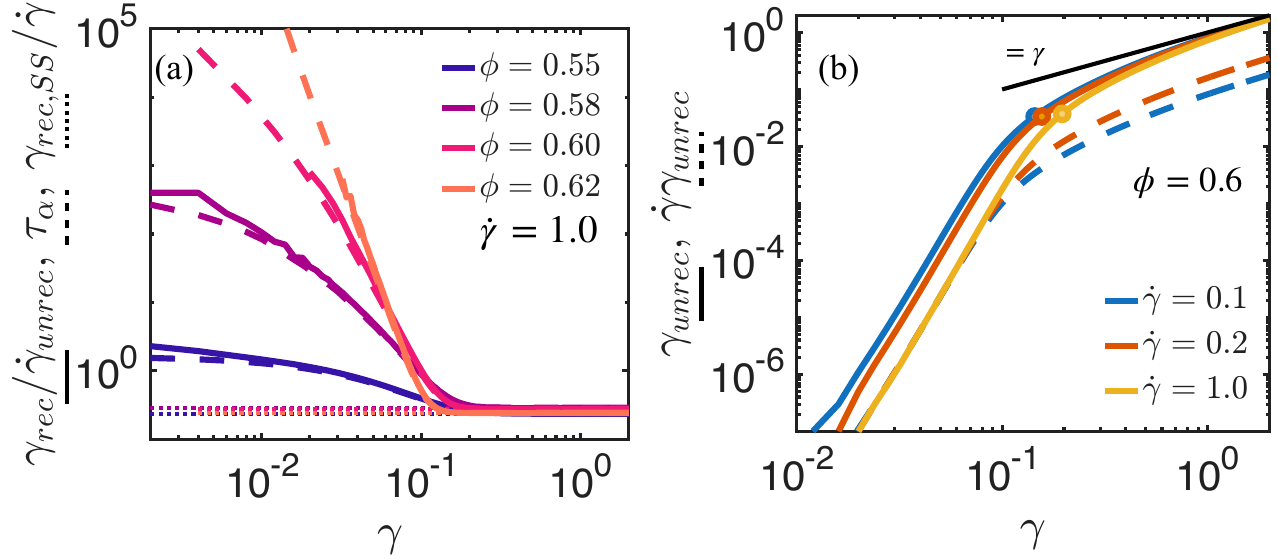}
  	\caption{ (a) The ratio $\gamma_{rec}/{\dot{\gamma}}_{unrec}$ (solid curves) is compared with the alpha relaxation time evolution (dashed curves) for different packing fractions, and with the steady state value of $\gamma_{rec,SS}/\dot{\gamma}$ (dotted lines). The predicted relation $\gamma_{rec}/{\dot{\gamma}}_{unrec}\sim\tau_\alpha$ holds in both the transient and steady-state regimes, and thus can be used to obtain transient relaxation time response. (b) Strain rate dependence of the unrecoverable strain growth (solid lines) as a function of applied total strain for $\phi=0.6$. In the limit of large strain, $ \gamma_{unrec}$ follows the total strain for all strain rates as it ultimately must. Dashed curves show a complete collapse of the  $\dot{\gamma}\gamma_{unrec}$ plots in the transient regime; see text for discussion. Solid circles in all plots mark the location of stress overshoot in the corresponding stress-strain curves.
  	} 
  	\label{fig7}
  \end{figure}
As a final application of our theoretical framework as applied to hard spheres, we explore how the accumulation of unrecoverable strain is linked with the time evolution of the structural relaxation time, a question typically not possible to directly probe experimentally with rheology in soft matter systems. The connection follows from first rewriting Eq. (\ref{GammaEff}) in terms of unrecoverable strain as: $\frac{d\gamma_{unrec}\left(t\right)}{dt}=\frac{\gamma_{rec}\left(t\right)}{\tau_\alpha\left(t\right)}$, resulting in $\gamma_{rec}/{\dot{\gamma}}_{unrec}=\tau_\alpha$. Numerically computed unrecoverable strain rates are shown in Fig.\ref{fig7}(a) and compared with the corresponding alpha relaxation time variation. Also shown are calculations of the quantity $\gamma_{rec,SS}{\dot{\gamma}}^{-1}$ (dotted lines) which corresponds to the theoretically predicted steady-state alpha time as discussed in Sec.\ref{Sec3b}. More generally, $\gamma_{rec}/{\dot{\gamma}}_{unrec}=\tau_\alpha$ links the acquired recoverable strain and unrecoverable strain rate to the structural relaxation time evolution, which under deformation can be greatly reduced from its typically unmeasurable quiescent value to Brownian-like accessible times. Such an alternative route to obtaining the alpha time evolution through recovery rheology can be highly valuable across diverse areas of soft matter research. Since the acquired unrecoverable strain is directly linked to the change in the fundamental alpha time under external deformation, this provides a foundation for explaining the observed quasi-universality of acquired unrecoverable strains across rheological protocols \cite{singh2021revisiting-26e}. 

The complementary calculations in Fig.\ref{fig7}(b) present the total strain evolution of the unrecoverable strain for three different applied shear rates. Systems deformed at smaller strain rates have more time to relax while being driven to the same total accumulated strain, and therefore acquire a larger  unrecoverable strain and yield at smaller strains. However, in the large strain nonequilibrium steady state limit, all curves converge to the total strain. Remarkably, the theory predicts that the transient unrecoverable strain growth collapses onto a single master curve when plotted as  $\dot{\gamma}\gamma_{unrec}$. This behavior directly follows from physics discussed above that unrecoverable‑strain growth is governed by the transient evolution of the structural relaxation time which evolves in the transient regime in a net strain-dependent manner. Mathematically, it can be  understood by rewriting Eq. (\ref{GammaEff}) as $\dot{\gamma}\gamma_{unrec}=\int_{0}^{\gamma}{\frac{\gamma_{rec}\left(\gamma,\dot{\gamma}\right)}{\tau_\alpha\left(\gamma,\dot{\gamma}\right)}d\gamma}$. In the transient regime prior to overshoot, the integral term on the right-hand side of this  relation becomes effectively independent of applied strain rate. This arises because $\gamma_{unrec}$ remains perturbatively small before the overshoot, implying $\gamma_{rec}\approx\gamma$, and therefore $\tau_\alpha$ becomes primarily a function of the total strain rather than the applied strain  rate.

\section{Beyond Hard Spheres: Soft Finite Range Repulsive Spheres}\label{Sec4}
\begin{figure}[t]
	\includegraphics[width=0.48\textwidth]{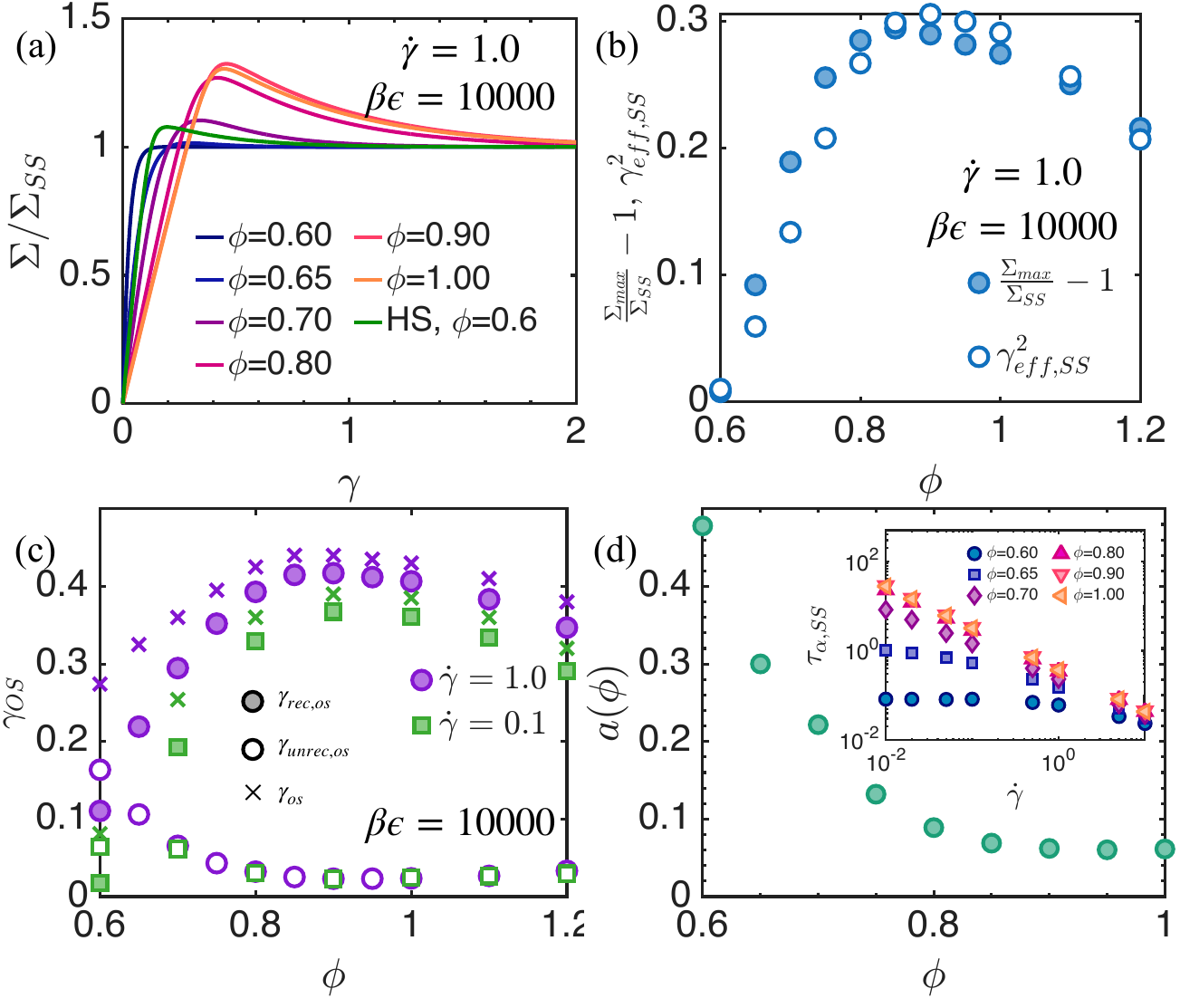}
	\caption{ (a) Stress evolution with strain for the soft particle system with $\phi\in[0.60-1.0]$ and $\beta\epsilon=10000$, compared to that of hard spheres (in green) for $\dot{\gamma}=1$. Stress is normalized by its steady-state value to highlight the enhanced overshoot for soft particles. (b) The quantitative relation ${\Sigma}_{max}/{\Sigma}_{SS}-1\ \sim{\gamma_{eff,SS}}^2=\left(\dot{\gamma}\tau_{\alpha_{SS}}\right)^2$ between the overshoot magnitude (filled points) and the steady-state deformation parameter (open points) is verified to hold (as expected) for the soft-particle system. Panel (c) depicts the overshoot strain $\gamma_{OS}$ as a function of $\phi$ for two different shear rates, along with its decomposition into unrecoverable ($\gamma_{unrec,OS}$) and recoverable ($\gamma_{rec,OS}$) components (d) Shear thinning plot is shown in the inset for different packing fractions, while the packing fraction variation of the exponent $a(\phi)$ per $\tau_{\alpha,SS}\sim{\dot{\gamma}}^{a-1}$ is presented in the main panel.  
	} 
	\label{fig8}
\end{figure}
A widely relevant and broad family of non-hard sphere soft matter systems are dense suspensions of tunably soft repulsive colloids or nanoparticles. A simple and popular model is described by non-divergent interparticle repulsive interactions of strictly finite spatial range. This model has been widely studied in simulations of the soft jamming crossover \cite{zhang2009thermal-64b}, and also the modelling of crosslinked microgels and related compact macromolecular based soft particles  \cite{vintha2026fluid-a76,kramb2011nonlinear-fe3}. Here, as a concrete example, we perform example calculations for such systems based on the classic Hertzian contact pair potential defined as:
 \begin{equation}\label{eq:HertzPot}
 	V(r)= 
 	\begin{cases}
 		\beta\frac{4\epsilon}{15}\left(1-\frac{r}{\sigma}\right)^\frac{5}{2},\ \ &r\le\sigma\\
 		0,\ \ &r>\sigma
 	\end{cases}
 \end{equation} 
In Eq (\ref{eq:HertzPot}), $r$ is the interparticle separation, $\sigma$ is a measure of particle diameter (repulsion onset distance), and $\frac{4\epsilon}{15}$ quantifies particle stiffness; the hard-sphere limit is recovered as $\epsilon\rightarrow\infty$. A measure of particle volume fraction is $\phi=\frac{\pi\rho\sigma^3}{6}$ by common convention. Equation (\ref{eq:HertzPot}) is considered realistic when soft particle deformation is small. The structure for a given $\phi$ and $\epsilon$ serves as input to the deformation-generalized ECNLE theory, which together with Eqs. (\ref{Maxwell}–\ref{GammaEff}) allows prediction of the rheological response under step-rate deformation. 

Figure \ref{fig8}(a) presents the  stress–strain results for a softness parameter $\beta\epsilon=10000$ and high packing fractions of $\phi\in[0.6-1.0]$. The behavior qualitatively follows the hard-sphere results of the main text, apart from the expected shift to higher $\phi$ and substantially larger overshoot magnitudes. Panel (b) plots the overshoot magnitude (solid markers) together with the steady‑state recoverable strain squared (open symbols). The results reinforce the relevance and generality of the quantitative connection established in Sec. \ref{Sec3c} for hard spheres. This enhanced overshoot magnitude is also physically expected since their softer, longer‑ranged repulsive interactions allow greater structural deformation, thereby leading to larger steady‑state cage distortion and, consequently, a larger stress overshoot. 

Fig.\ref{fig8}(c) shows that the total and recoverable overshoot strain are predicted to exhibit a non-monotonic dependence on packing fraction, qualitatively similar to the hard sphere system, but with the peak shifted to larger $\phi$ for soft particles. The unrecoverable overshoot strain, however,  monotonically decreases with  $\phi$ and, as established for hard spheres in Sec. \ref{Sec3d}, remains invariant with respect to shear rate. These results support the generality of our theoretical result that the stress overshoot (a metric for a continuous solid–fluid transition or yield point) corresponds to the system accumulating a characteristic, strain rate‑independent unrecoverable strain. The strain‑rate‑dependent mechanistic definition of solidity, based on separating the overshoot strain into recoverable and unrecoverable components, becomes even more significant for this dense soft‑particle system.
 
Finally, Fig.\ref{fig8}(d) (inset) shows the shear-thinning behaviour of the steady‑state relaxation time as a function of imposed shear rate. The apparent high-rate power-law exponent deviates further from $-1$ than for the hard sphere systems. This is consistent with the connection developed in Sec.
\ref{Sec3b} linking stronger steady‑state cage deformation with increasing $\dot{\gamma}$ in soft particles. The effective thinning exponent $a(\phi)$ is shown in Fig.\ref{fig8}(d), which decreases with increasing packing fraction and approaches a very low value at the soft jamming crossover. A value of exactly zero, as occurs at RCP for hard spheres that literally jam, is not expected for soft particles, which can overlap.

Overall, our results for this soft particle system confirms our expectation  that the basic interconnections predicted in Secs.\ref{Sec3b}, \ref{Sec3c}, and \ref{Sec3d} for hard spheres hold more broadly. This buttresses our argument that the results in Sec.\ref{Sec3e} are purely physically-motivated mathematical consequences of the theory construction that transcend the precise interparticle interaction potential.

  \section{High Level Comparisons with the KDR Model}\label{Sec5}
  We now briefly discuss connections of our microscopic theoretical approach to the phenomenological KDR model. Both descriptions are fundamentally built on the recovery‑rheology framework and the separation of strain into its recoverable and unrecoverable components as proposed by Rogers and coworkers  \cite{shi2023benefits-879,singh2021revisiting-26e,kamani2021unification-578,donley2020elucidating-c91,keane2025universal-80b}. Of course, precise connections for all aspects are impossible to make, given the non-microscopic, continuum model nature of the KDR approach. Moreover, recall that a major theme in our work above is to propose a microscopic formulation of recoverable strain in terms of the microscopic structure under deformation within a wavevector advection framework that retains predictive power.
  
  The KDR formulation employs a Maxwell-type equation for stress generation similar to Eq.(\ref{Maxwell}), but with several key differences. First, the KDR step‑rate equation includes an additional short-time viscous stress‑reduction term involving a quiescent shear viscosity and relaxation time of the form 
  $-\frac{\eta_s}{\tau_\alpha}\dot{\gamma}$. This term is small relative to the dominant contributions in step-rate deformation measurements, and has minimal influence on the overall stress evolution in the deeply metastable states of interest to us. It was thus purposefully not included in our formulation, 
  but could be. 
  
  Second, and far more consequential with regard to the basic physics, the KDR model assumes that both the elastic modulus $G^\prime$ and the relaxation time $\tau_\alpha(\dot{\gamma})$ are time (or total strain) \textit{independent}. The modulus is fixed at its experimentally measured quiescent value, whereas the relaxation time is taken by ansatz directly from the steady‑state flow curve (stress versus shear rate) modelled mathematically using the empirical Hershel-Buckley (HB) relation with fit parameters. Thus, the relaxation time is a function of shear rate alone under \textit{all} transient and steady state conditions, and not coupled with the instantaneous microstructural state or accumulated strain. None of these simplifications are adopted in our theoretical approach.
  
  From our perspective, the above second set of issues is the key physical shortcoming of the KDR model. As shown in Sec.\ref{Sec3c}, from our theoretical perspective the deformation-induced softening of the elastic modulus is the fundamental origin of the stress overshoot. In contrast, adoption of the steady-state strongly reduced (shear-thinned) relaxation times as a surrogate for dynamics in the transient regime will necessarily overpredict the plastic effects in the transition regime from elastic-like to viscous-like rheological response. Indeed, Rogers et. al. have shown \cite{kamani2024brittle-e9b} that the KDR model in this form cannot predict a correct step-rate transient response that exhibits a stress overshoot, consistent with the physical origin of the transient overshoot behavior we have proposed. However, interestingly, the simplified treatment of KDR has been shown to produce a correct transient response for the very different large amplitude oscillatory shear (LAOS) rheological experiments \cite{donley2020elucidating-c91}. This finding will be addressed within our theoretical framework in a future work. 
  
  Recently, an interesting attempt to phenomenologically incorporate the missing time dependence of the relaxation time in the KDR framework has been proposed using the recovery rheology parameters and an effective strain rate of the form ${\dot{\gamma}}^\prime(t)=\frac{{\dot{\gamma}}_{rec}(t)}{Bt}+{\dot{\gamma}}_{unrec}(t)$ in the calculation of $\tau_\alpha\left(\dot{\gamma}^\prime\right)$. Here, a positive ``brittility” parameter $Bt>1$ has been introduced \cite{kamani2024brittle-e9b}, which in practice is empirically chosen to fit data on a system-by-system basis. Successful applications to experiments have been demonstrated  \cite{kamani2024brittle-e9b}. Conceptually, within our theoretical perspective, a factor such as Bt does \textit{not} enter. We view its use to realize a modified strain rate as serving to mimic the missing time dependence of the relaxation time. Specifically, at small times when ${\dot{\gamma}}_{unrec}\approx0$, the net effective strain rate is reduced by a factor of $Bt$. This results in a large $\tau_\alpha\left(t\rightarrow0\right)$ and thus the desired small plastic effects, while in the steady-state when ${\dot{\gamma}}_{unrec}\rightarrow\dot{\gamma}$ the steady state relaxation time is recovered. Thus, the introduction of $Bt\neq1$ leads to relaxation time evolution from larger $\tau_\alpha\left(\frac{\dot{\gamma}}{Bt}\right)$ at initial times to a smaller $\tau_\alpha\left(\dot{\gamma}\right)$ in the steady state, with the extra input of ${\dot{\gamma}}_{unrec}\left(t\right)$ from separate recovery experiments or other model approximations. However, the implementation of elastic modulus softening is still missing, although it has some experimental justification for the LAOS rheological measurement  \cite{kamani2023understanding-4db}. 
  
  To connect the KDR model to our theoretical approach at a high level, the stress evolution is obtained by a similar Maxwell-like model, with the transient relaxation time interpreted from recovery rheology (for $Bt\neq1$) and steady-state flow curve. Thus, in an approximate sense, the KDR model reverse engineers the structural evolution of Eq.(\ref{GammaEff}) via recovery rheology to obtain the transient time-dependent relaxation time required as input to predict stress evolution. In Sec.\ref{Sec3e}, we suggested what we believe is a more accurate and physically motivated alternative. Specifically, the ratio $\frac{\gamma_{rec}}{{\dot{\gamma}}_{unrec}}$  provides a measure (conceptually superior to empirical HB flow curve based prediction) of the relaxation time, given the recoverable-unrecoverable strain decomposition is available.  
  
  \section{Discussion and Future Directions}
    We have presented and quantitatively applied in detail a predictive statistical-mechanical theory for recovery rheology, rooted in quiescent and nonequilibrium-activated glass physics, that provides connections between microscopic quantities and phenomena and macroscopic rheological measurements. A core new idea is that \textit{recoverable} strain is determined by a well-defined microscopic measure of local structural deformation and mechanical memory, with the unrecoverable part representing plastic loss due to irreversible thermally activated processes in Brownian fluids and suspensions. In the nonequilibrium steady state flow regime, the variation of recoverable strain or microstructural deformation with shear rate directly causes deviations from the ideal shear thinning behavior of the structural alpha and stress relaxation times. In step-rate shear measurements, the stress overshoot feature results from structural over-deformation, surpassing the limit of inter-particle forces to store energy elasticity. This results in the prediction that the shear-rate dependent steady-state recoverable strain is quantitatively related to the transient stress overshoot magnitude. These previously unknown connections represent predictions that are testable in experiment and simulation. 
    
    The overshoot strain or yield point is typically qualitatively viewed as indicative of a continuous solid-to-liquid crossover or transition. Our theory predicts it corresponds to the acquisition of a shear-rate-independent unrecoverable strain, thereby endowing this important feature with a precise physical meaning. On the other hand, the rate at which the unrecoverable strain is acquired directly relates to relaxation\ time evolution under deformation. These connections help microscopically address the central physics question of material evolution under different deformations. This includes nonlinear oscillatory strain and stress and constant stress creep responses, which are of both fundamental nonequilibrium physics and high material applications importance. As a potentially integrative future direction, we believe that further comparison of our theoretical time-dependent microscopic rheology framework to the highly useful phenomenological KDR model  \cite{kamani2021unification-578}  holds the promise to provide deeper physical insights into the origin and rational tunability of its various fitting parameters.
  
    Finally, although we have illustrated the predictive power of the new theory for the foundational metastable glass forming hard sphere suspension, the basic ideas are general for all spherical particle systems across diverse chemistries as encoded in the interparticle pair potential. Initial applications in this direction have been presented for a simple model of soft repulsive colloids, and qualitative similarities and quantitative differences in predicted behavior compared to the hard sphere system have been illustrated. Applications of our approach to other classes of dense soft colloidal matter, including charged colloids and attractive particles that can form gels and attractive glasses, are also possible.
    
\noindent{\bf Acknowledgements:}
The authors acknowledge support from the Army Research Office via a MURI grant with Contract No. W911NF-21-0146, and stimulating discussions with Simon Rogers.
\vskip +0.1in
\noindent{\bf Data availability:} The data that support the findings of this article are openly available \cite{Data}.
\bibliography{StepRateOverShoot}

\begin{thebibliography}{55}
\bibitem [{Dat()}]{Data}%
  \BibitemOpen
  \href@noop {} {}\bibinfo {note} {The data is openly available at
  10.6084/m9.figshare.31510903"}\BibitemShut {NoStop}%
\end{thebibliography}

\appendix
\section{Background Theoretical Elements}\label{ApdA}
\subsection{Short-time in-cage relaxation process time scale, $\mathbf{\tau}_\mathbf{s}$}\label{ApdA1}
The explicit expression for $\tau_s$ of hard sphere fluids is (for details see Ref.  \cite{schweizer2005derivation-c99}). 
\begin{equation}\label{Taus}
	\begin{split}
	\tau_s\equiv&\ \tau_0g(\sigma)\left[1+\frac{\sigma^3}{36\pi\phi}\int_{0}^{\infty}{dk\frac{k^2{(S(k\sigma)-1)}^2}{S(k\sigma)+b(k\sigma)}}\right],\\ &b^{-1}(\sigma k)\equiv1-j_0(k\sigma)+2j_2(k\sigma)
	\end{split}
\end{equation}
Here, $j_n(x)$ is the spherical Bessel function of order $n$, and $\tau_0\equiv\frac{\zeta_{SE}\sigma^2}{k_BT}$ is the ``bare” elementary time written in terms of the Stokes-Einstein friction constant, $\zeta_{SE}$, and the contact value of the pair correlation function, $g(\sigma)$.

\subsection{Static structure factor and the RCP limit}\label{ApdA2}
The structure factor for hard sphere fluids is obtained from highly accurate integral equation theory with the modified-Verlet (MV) closure, which has been shown to be quantitatively good in the deeply metastable regime (up to $\phi=0.585$) by comparison to crystal-avoiding simulations  \cite{zhou2020integral-c4b}. Although the RCP packing fraction predicted by the MV closure is too large $\phi_{RCP,MV}\approx0.776$ (true of all integral equation theories) compared to the monodisperse smooth hard sphere value of $\phi_{RCP}\approx0.644$, it is a major improvement over the classic Percus–Yevick (PY) \cite{schweizer2005derivation-c99} prediction of  $\phi_{RCP,PY}=1$, and correctly captures several jamming-like laws for thermodynamic properties and the pair correlation function  \cite{chaki2024theoretical-92e,athanasiou2025probing-0e7}.  

\subsection{ECNLE theory extension}\label{ApdA3}
NLE theory predicts the particle jump length required to cross the barrier and escape its cage is sufficiently large that a small collective elastic expansion of the cage is required resulting in a spatially nonlocal contribution to the alpha relaxation event (see Fig.\ref{fig1}(a)). Such a non-local extension of the NLE approach defines the Elastically Collective NLE (ECNLE) theory  \cite{mirigian2014elastically-4c6} via the addition of an elastic barrier to the local cage one; both barriers are causally related via the underlying dynamic free energy. The elastic barrier is calculated within the Einstein glass model for the localized particles outside the cage as $F_{el}=4\pi\int_{r_{cage}}^{\infty}r^2\rho g\left(r\right)\left(\frac{1}{2}K_0u\left(r\right)^2\right)dr$, where $K_0$ is the harmonic spring constant of $F_{dyn}$ at its minima, and $u\left(r\right)$ is the isotropic displacement field for particles outside the cage at a distance $r$ from the cage center (see Fig.\ref{fig1}(c)). The displacement field is modeled in the spirit of continuum linear elasticity  \cite{dyre2006elastic-c3c} as $u\left(r\right)=\Delta r_{eff}\left(\frac{r_{cage}}{r}\right)^2$ for $r\geq r_{cage}$, where the cage radius $r_{cage}$ is identified as the first minimum of $g\left(r\right)$, and $\Delta r_{eff}$ quantifies the effective cage expansion. The latter length scale is obtained by performing a microscopic analysis of the mean extent to which cage scale hopping results in a particle displacement larger than the cage size in terms of the microscopic jump distance $\Delta r=r_B-r_L$ to yield,  \cite{mirigian2014elastically-4c6}
\begin{equation}
	 \Delta r_{eff}\approx\frac{3}{r_{cage}^3}\left(\frac{r_{cage}^2\Delta r^2}{32}-\frac{r_{cage}\Delta r^3}{192}+\frac{\Delta r^4}{3072}\right).
\end{equation}
The large amplitude activated Brownian hopping dynamics including the collective elasticity contribution then defines ECNLE theory  \cite{mirigian2014elastically-4c6}. 

\section{Effect of structure and stress relaxation mismatch factor}\label{ApdB}
\begin{figure}
	\includegraphics[width=0.45\textwidth]{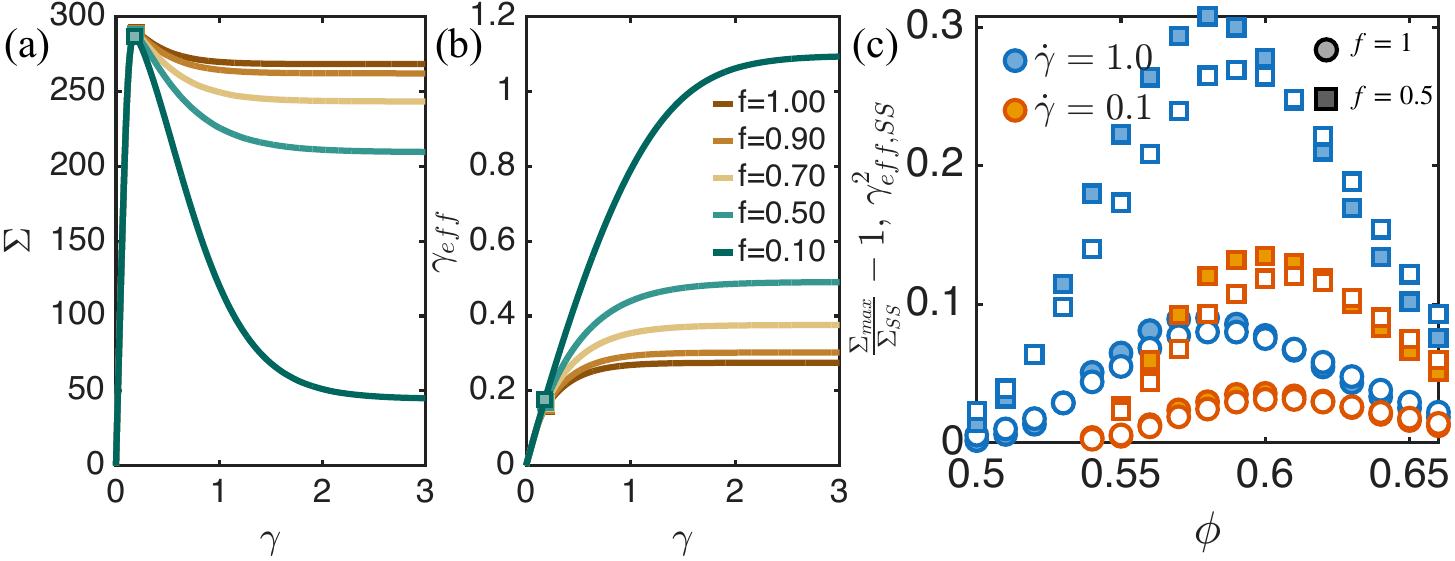}
	\caption{ Effect of the mismatch factor between the stress and structural relaxation times on (a) stress-strain evolution and (b) structure evolution. For $f<1$ the stress overshoot and steady state cage deformation are larger. The quantitative relation between them established in the main text,  $\mathrm{\Sigma}_{max}/\mathrm{\Sigma}_{SS}-1\ \sim{\gamma_{eff,SS}}^2=\left(\dot{\gamma}\tau_{\alpha_{SS}}\right)^2$, is tested in (c) for $f=0.5$. The filled symbols show the stress overshoot magnitude as a function of $\phi$ for two different shear rates, along with its comparison to $\gamma_{eff,SS}^2$ in the open symbols. Two different colors indicate different applied shear rates, while different symbols represent two different structure-to-stress relaxation time ratios.
	} 
	\label{fig9}
\end{figure}
In Eq. (\ref{GammaEff}), one can introduce a \textit{constant numerical} factor $f$ to model a possible modest nonuniversal mismatch (well known to occur in quiescent glass-forming liquids  \cite{denisov2013resolving-853}) between the rate of stress and structure relaxation, $\tau_\alpha^{structure}=\tau_\alpha^{stress}/f$, as, 
\begin{equation}
	\frac{d\gamma_{eff}\left(t\right)}{dt}=-f\frac{\gamma_{eff}\left(t\right)}{\tau_\alpha\left[\Sigma\left(t\right),\gamma_{eff}\left(t\right)\right]}+\dot{\gamma}. 
\end{equation}
Calculations in the main text adopted the minimalist choice of $f=1$.  However, structural relaxation may involve quantitatively larger length and time scales than stress relaxation for highly activated dynamics systems. This motivates a brief study $f<1$ for the hard sphere fluid.  

Fig. \ref{fig9}(a) presents calculations of the stress-strain curves for an order of magnitude variation of this mismatch factor. With decreasing $f$, the internal structure relaxes more slowly, leading to larger cage deformation (Fig. \ref{fig9}(b)) in the steady state, and hence a larger overshoot, as explained in the main text. The increased amplitude of the overshoot is clearly visible, while the theoretically predicted connection discussed in the main text of $\Sigma_{max}/\Sigma_{SS}-1\ \sim{\gamma_{eff,SS}}^2=\left(\dot{\gamma}\tau_{\alpha,SS}\right)^2$ is tested and again well confirmed in Fig. \ref{fig9}(c) for $f=0.5$. For context, if the structural relaxation is 10 times slower (half) than the stress relaxation, the overshoot magnitude increases from around $0.1$ for $f=1$ to $5.0$ for $f=0.1$ ($0.3$ for $f=0.5$).

\end{document}